# A Novel Methodology of Visualizing Orthorhombic Phase Uniformity in Ferroelectric Hf$_{0.5}$Zr$_{0.5}$O$_2$ Devices Using Piezoresponse Force Microscopy

Wei-Cheng Peng, Hsien-Yang Liu, Cheng-Yu Yu, Artur Useinov, Tian-Li Wu, *member IEEE*

*Abstract* Ferroelectric Hf$_{0.5}$Zr$_{0.5}$O$_2$ (HZO) thin films are promising for next-generation memory and logic devices due to their CMOS compatibility and scalability. The spatial uniformity of the orthorhombic (O) phase is crucial for optimizing ferroelectric properties like remnant polarization. This work introduces a novel piezoresponse force microscopy (PFM) approach for 2D mapping of O-phase uniformity in HZO films (5 nm, 9 nm, and 20 nm), further quantifing O-phase distribution by distinguishing polarized O-phase regions from non-polarized tetragonal/monoclinic (T/M) phases. Our results reveal that the 9 nm film exhibits the most uniform O-phase and highest remnant polarization. This PFM-based method enables comprehensive phase characterization without requiring complicated facilities, broadening access to phase analysis and advancing ferroelectric thin-film research for memory and logic applications.

*Index Terms*—**Hf$_{0.5}$Zr$_{0.5}$O$_2$ Ferroelectric, Piezoresponse Force Microscopy.**

## I. INTRODUCTION

Ferroelectric HZO thin films represent a transformative advancement for next-generation memory and logic devices, offering substantial benefits due to their compatibility with CMOS processes and scalability [1]-[6]. Ferroelectric materials exhibit spontaneous electric polarization that can be reversed by an applied electric field, making them highly suitable for applications such as non-volatile memory, low-power logic, and energy-efficient electronics.

Accurate phase characterization is crucial for optimizing the performance of HZO-based devices. Grazing Incidence X-Ray Diffraction (GIXRD) is commonly used to determine the presence of orthorhombic (O-), tetragonal (T-), and monoclinic (M-) phases in HZO films [7]-[8]. However, the typical GIXRD can only analyze the phase composition in one specific cross section. Two-dimensional (2D) phase mapping is preferred to understand the phase distribution in the whole area, further providing the deeper insights into phase uniformity. The recent literature reported 2D phase mapping in Hf$_{1-x}$Zr$_x$O$_2$ using synchrotron X-ray technique [9]. However, the synchrotron X-ray technique requires a complicated facility capable of generating extreme high-energy X-rays for X-ray absorption spectroscopy. In contrast, PFM technique is simple as well as often utilizing to characterize ferroelectric properties of the thin films due to piezo response effect [11]-[12]. The previous studies reported that PFM can be used to characterize ferroelectric performance and switching behavior [11]-[15], However, 2D phase mapping based on the PFM measurements has not been reported yet.

In this study, we present a novel methodology for visualizing the O-phase uniformity as the two-dimensional (2D) phase mapping in HZO devices through piezoresponse force microscopy (PFM). For the first time, PFM is employed to quantify the O-phase percentage and its spatial distribution in HZO thin films with varying thicknesses (5 nm, 9 nm, and 20 nm). We successfully analyzed 2D O-phased percentage and uniformity in HZO films with different thicknesses of 5 nm, 9 nm, and 20 nm, further correlating 2D O-phased percentage with 2P$_r$ values in HZO samples.

## II. DEVICE SCHEMATICS AND CHARACTERISTICS

The schematic illustrating of metal/insulator (HZO)/metal (MIM) process flow and sample structure are presented in Fig.1(a) and Fig.1(b). The devices were fabricated on a p-type Si wafer. The deposited thickness of 75 nm and 5 nm of the TiN electrodes were deposited by physical vapor deposition (PVD) and by atomic layer deposition (ALD), respectively. The bottom 5 nm TiN, 5 nm (or 9 nm/ 20 mn) HZO, and the top 5 nm TiN layers are deposited by ALD in the same chamber to avoid the formation of the interfacial layer [16]. The ferroelectric layer was crystalized by rapid thermal annealing (RTA) at 600℃ for 30 seconds. Subsequently, the characteristics of the ferroelectric layer were measure by Keysight B1500A with WGFMU module. PFM measurements were performed using a Dimension ICON atomic force microscope (Bruker Technologies) in a contact mode. Contact mode was conducted with a writing rate of 1 Hz

Manuscript received XXX, 2025; revised XXX, 2025; accepted XXX, 2025. Date of publication XXX, 2025; date of current version XXX, 2025. This work was supported in part by the "Advanced Semiconductor Technology Research Center" from the Featured Areas Research Center Program within the framework of the Higher Education Sprout Project by the Ministry of Education (MOE) in Taiwan. The authors thank Instrumentation Center of National Tsing Hua University for the support of the Scanning Probe Microscope (SPM, PFM Mode) for PFM measurements. The review of this letter was arranged by Editor XXX. (*Corresponding author: Tian-Li Wu*)
H.-Y. Liu and T.-L. Wu are with Institute of Electronics, National Yang Ming Chiao Tung University, 30010 Hsinchu, Taiwan. (e-mail: tlwu@nycu.edu.tw)
W.-C. Peng, C.-Y. Yu, and A. Useinov are with International College of Semiconductor Technology, National Yang Ming Chiao Tung University, 30010 Hsinchu, Taiwan, and a reading rate of 0.898 Hz. The writing voltage was normalized to -4 MV/cm at a reading voltage of 0.8 – 1.0 V for all samples.

The GIXRD analysis, as illustrated in Fig. 1(c), reveals a characteristic peak at 30.5° in the annealed samples, indicating a mixed O/T phase in all HZO samples. Additionally, thickness-dependent O/T-phase signal intensity is well observed. The highest O/T-phase signal is displayed for 9 nm sample followed by a slightly lower intensity for 20 nm one, while 5 nm sample exhibited the weakest signal at 30.5°. The 20 nm sample notably demonstrated a significant M-phase signal response.

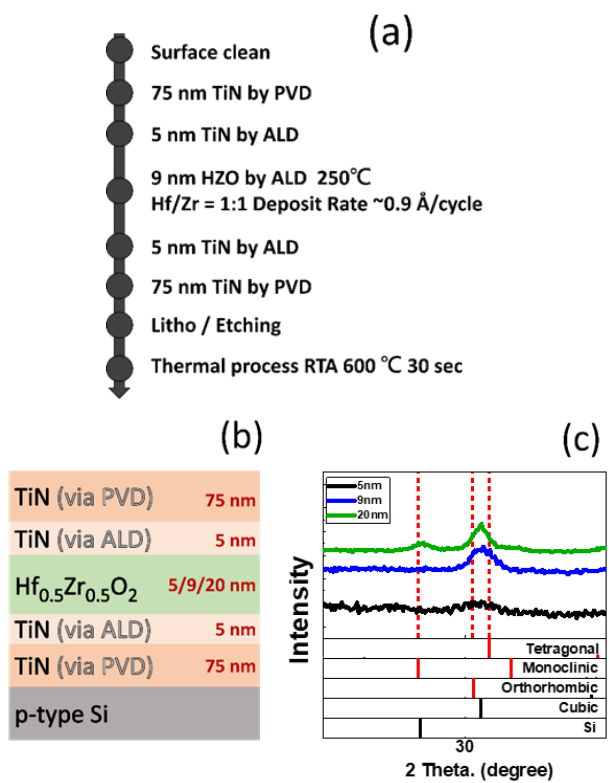

Fig. 1 (a) Fabrication flow for HZO-based MIM structure, (b) Schematic drawing of the MIM system, and (c) GIXRD results for 5 nm, 9 nm, and 20 nm of MIM samples.

These crystallographic observations correlate well with the polarization-voltage (P-V) characteristics in Fig. 2 (a). The 9 nm and 20 nm samples exhibit higher $P_r$ values compared to the 5 nm sample when they are measured under an applied electric field of 4MV/cm. The influence of the M-phase on the ferroelectric properties is particularly evident in the 20 nm sample, which shows a lower $2P_r$ value than the 9 nm sample (Fig. 2(b)).

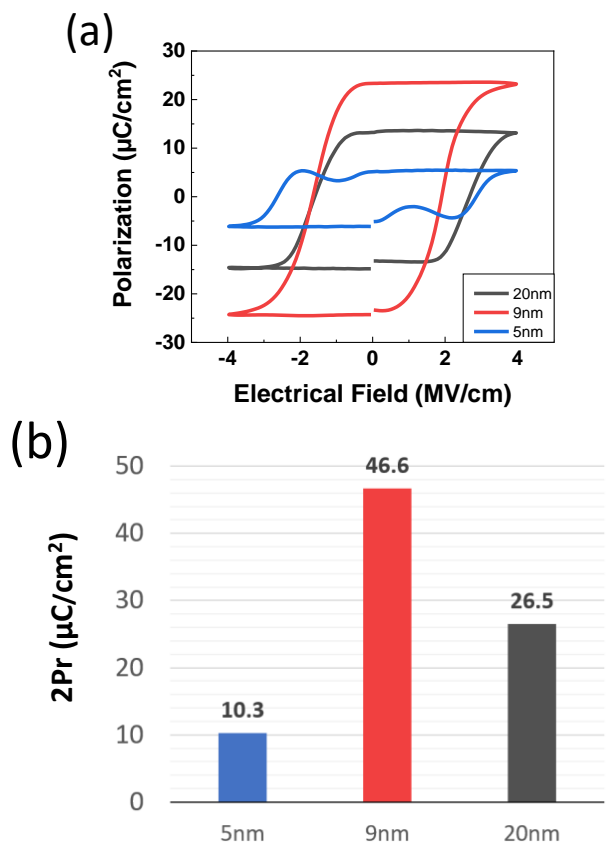

Fig. 2 (a) P-V loop under 4 MV/cm electric field in the fresh state of the device and (b) 2Pr comparison between samples.

### III. METHODOLOGY AND RESULTS

Despite the potential antiferroelectric behavior observed for the 5 nm HZO device [17], our study focuses on comparing the $P_r$ values. We can effectively disregard any potential influences of antiferroelectric effects in our samples by ensuring that all our devices are subjected by sufficient electric fields during the polarization process.

To analyze the O-phase crystal structure, we demonstrate the O-phase and non-O-phase 2D mapping methodology using the PFM. Fig. 3 illustrates the methodology flow. First, we conducted P-V measurements using the Positive Up Negative Down (PUND) to determine $2P_r$ and coercive field Ec (Fig. 3.1). A fresh HZO sample exhibits a random distribution of O-, T-, and M- phases (Fig.3.2-1). Before the PFM measurement, the device is biased with a writing voltage ($V_{write}$) to uniformly polarize the O-phase crystal domains by applying a high electric field during the writing process can ensure that all polarizable regions are fully switched and aligning them all O-phase crystals in the same direction (Fig. 3.2-2). Then, the reading voltage ($V_{read}$) below the coercive voltage $V_c$ is conducted to ensure a pure O-phase polarization response and phase signal, which can be used as the piezoresponse amplitude and phase mapping to identify polarized area of the O-phase (Fig. 3.2-3). As a result, the phase signals recorded during $V_{read}$ scanning can reflect the characteristics of the original O-phase crystal domain. At the moment of $V_{write}$ = 0, the P can be characterized by the correlation with the O-phase signal.

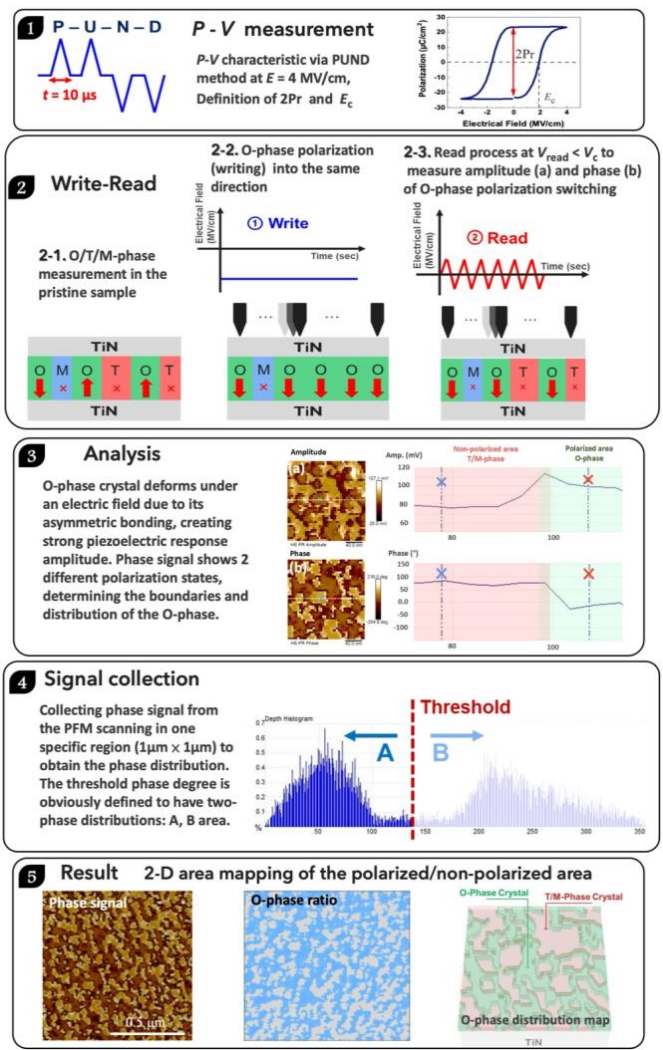

Fig. 3. Methodology for O-phase and non O-phase 2D mapping using the PFM measurement.

For further evaluation of the polarized O-phase and non-polarized T/M-phase regions, the piezoresponse amplitude and the phase analysis from the PFM are compared (Fig. 3.3). The O-phase domains, characterized by their asymmetric bonding, and exhibiting a strong piezoelectric response, lead to the obvious valuable piezoresponse. Furthermore, the phase signal reveals the distinct polarization states that mainly come from the different polarization directions. Therefore, combining the piezoresponse amplitude and phase analysis, it is possible to identify the polarized area caused by O-phase, since the O-phase has a strong amplitude signal and a clear distinct phase state. Fig. 4(a) provides a close-up of the 9 nm sample with the piezoresponse amplitude. The blue marker denotes a region with a weak amplitude response, while the red marker indicates a distinct signal response. Fig. 4(b) illustrates the phase signal at the same locations shown in Fig.6 (a), with the red and blue markers representing different polarization states. We can confirm the area with high piezoresponse amplitude, indicated as the red marker in Fig. 6(b), and the distinct polarization state in the O-phase area.

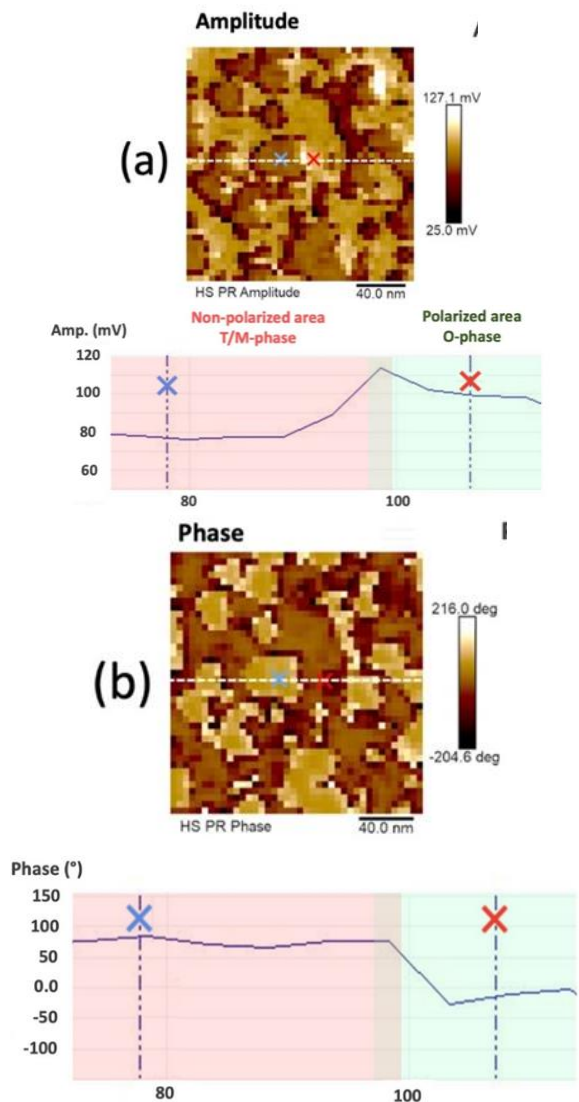

Fig. 4. (a) piezoresponse amplitude in O-phase crystal and T/M-phase crystal and (b) the phase differences.

After identifying the polarization region caused by O-phase areas, phase distribution in one specific region can be obtained from the PFM scanning (Fig. 3.4). The threshold phase degree is defined to identify two-phase distributions: area A for polarized and area B for non-polarized mapping. Note, that the slight shift of threshold phase degree in the valley region between the two distributions have the limited impacts on estimation of O-phase ratio since the O-phase ratio is mainly determined by the integration of the main obvious distribution peak in area A.

The final step (Fig. 3.5) involves quantifying the O-phase ratio via the phase threshold between polarized and non-polarized domains (see the red dashed line in Fig. 3.4). This threshold is typically located in the valley between the two peaks, distinguishing the initial O-phase from others phase distributions in 2D mapping.

Fig. 5 illustrates the piezoresponse amplitude in O-phase versus T/M-phase crystals. The O-phase crystal displays a strong piezoelectric response, indicating significant expansion/contraction along the vertical axis due to its asymmetric bonding. The T/M-phase structure shows a weak piezoelectric response. The alignment of polarization direction with the same direction of the applied electric field results in a 0° phase difference (Fig. 5(a)). When the polarization direction opposes the applied electric field, a 180° phase difference occurs. This 180° difference arises because oppositely polarized crystals respond differently to the same electric field [18]. On the other hand, regions with minimal piezoelectric response typical to T/M-phases, where signals are nearly indistinguishable from background noise (Fig. 5(b)).

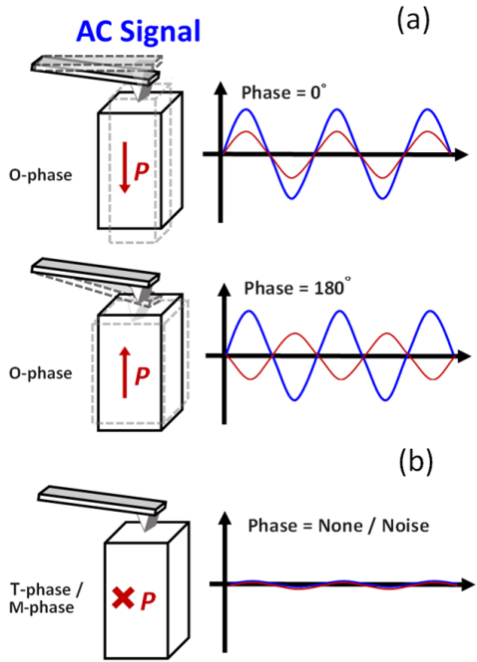

Fig. 5. Schematic illustration of (a) piezoresponse amplitude in O-phase crystal and its phase difference for P and E-field alignment; (b) T/M-phase crystal with minimal piezoelectric response.

The PFM measurements were conducted using high-frequency piezoresponse force microscopy (PFM) near the contact resonance frequency with the scanning area 500nm*500nm. The operating frequency range was 150 kHz to 290 kHz. A cantilever with a spring constant of 2.8 N/m was utilized. The lock-in amplifier settings included a lock-in bandwidth of 0.75 kHz, a vertical 16x gain (enabled), and a drive amplitude adjusted up to 2000 mV to prevent material damage. Fig. 6(a)(b) and (c) presents the piezoresponse amplitude and phase results for the 5 nm, 9 nm, and 20 nm HZO samples. The piezoresponse amplitude is increased as HZO thickness increases, indicating stronger O-phase crystal responses. Furthermore, Fig. 6(d)(e)(f) illustrates the phase response that can be used to distinguish the polarized O-phase and non-polarized T/M-phase regions.

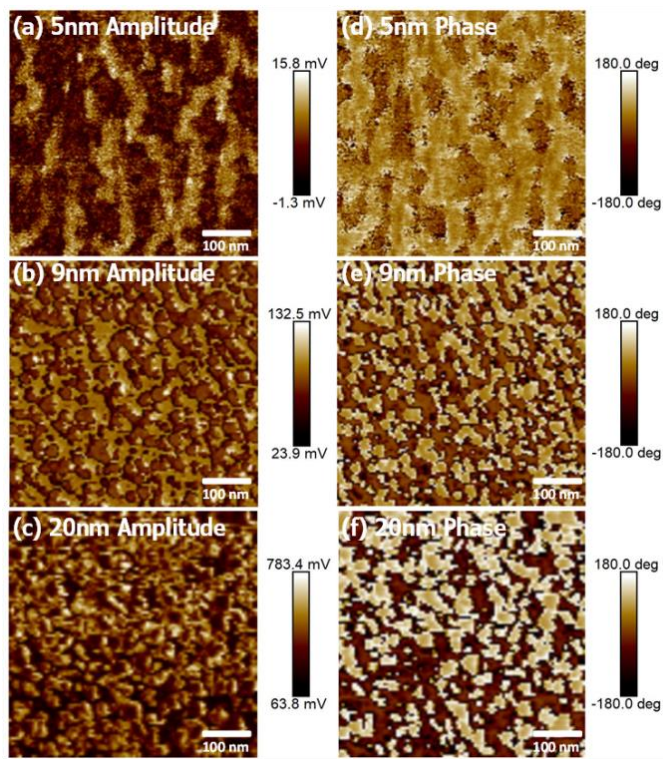

Fig. 6. PFM result for the zoomed area in 9 nm sample: piezoresponse amplitude (a-c), and phase signal results in (d-f).

Based on the proposed methodology, the 2D O-phase uniformities are analyzed in the sample with different HZO thicknesses. The analysis of the O-phase ratio in the HZO samples with different thicknesses are summarized in Fig. 7. In Fig. 7(a), the O-phase ratio of the 5 nm device is only 34.45%. Furthermore, the O-phase ratio is significantly increased to 56.95% for the 9 nm sample (Fig. 7(b)). Subsequently, Fig. 7(c) illustrates a slight decrease in the O-phase ratio to 47.59% for the 20 nm device. Based on the mapping of the O-phase ratio and its uniformity, the results demonstrate strong consistency with other research [19].

The comparison between the O-phase ratio (Fig. 7) and $2P_r$ of the samples (Fig. 2) with different HZO thicknesses indicates a strong correlation between $2P_r$ values and the O-phase ratios. Comparing the samples with 5 nm and 9 nm HZO thicknesses, the obvious decrease of $2P_r$ can be mainly attributed to the reduction of the O-phase ratio for 5 nm sample. On the other hand, increasing the HZO thickness to 20 nm reduces the $2P_r$ as well as the O-phase ratio, which is consistent with literature [17], reported that HZO with thickness greater than 10nm can increase M-phase ratio and decrease $2P_r$ value.

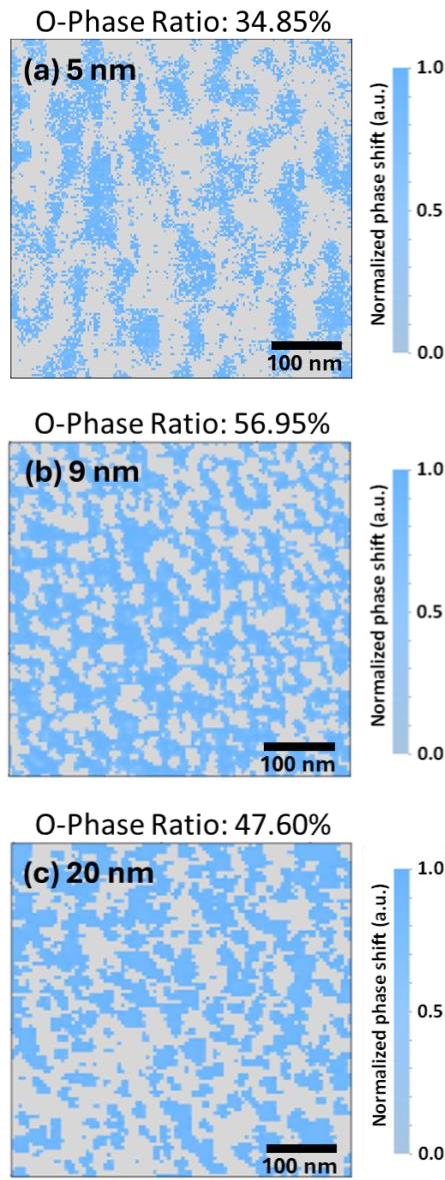

Fig. 7. O-phase ratio for 5 nm (a), 9 nm (b), and 20 nm (c) HZO layer thickness.

## IV. CONCLUSION

In this study, we successfully demonstrated a method to create two-dimensional maps of the orthorhombic (O-phase) distribution and its uniformity in HZO films with different thicknesses (5 nm, 9 nm, and 20 nm). By using PFM, we are able to find correlations between $P_r$ and amplitude with the O-phase distribution, showing a clear relationship of the ferroelectric performance with the proportions of the O-phase domains area.

Our results show that the 9 nm HZO sample had the highest proportion of the orthorhombic phase, which was associated with the highest $P_r$. This emphasizes the importance of phase uniformity for optimizing the ferroelectric properties of HZO films, especially for achieving higher $P_r$ values. Additionally, our work highlights that PFM can be used as an effective tool for phase mapping without requiring access to the complicated facilities, making it more practical and accessible for a wider range of research. The PFM methodology of the 2D mapping that we developed offers a novel approach for a phase distribution studying of ferroelectric thin films, which is essential for speeding up the advancement of the next-generation non-volatile memory and low-power ferroelectric devices.